\documentclass[]{article} 
\usepackage{xspace}
\usepackage{tabulary}
\usepackage{colortbl}
\usepackage{microtype}
\usepackage{graphicx}
\usepackage{siunitx}
\usepackage{pifont}
\usepackage{glossaries}
\usepackage{subcaption}

\definecolor{darkgreen}{RGB}{0,100,0}

\newcommand{\appName}{\textit{HandyLabel}\xspace}
\newcommand{\labelStudio}{\textit{Label Studio}\xspace}
\makeglossaries
\glsdisablehyper
\newacronym{ai}{AI}{Artificial Intelligence}
\newacronym{llms}{LLMs}{Large Language Models}
\newacronym{igms}{IGMs}{Image Generation Models}

\usepackage{cite}
\usepackage{amsmath,amssymb,amsfonts}
\usepackage{algorithmic}
\usepackage{graphicx}
\usepackage{subcaption}
\usepackage{textcomp}
\usepackage{xcolor}
\usepackage{multirow}
\usepackage{adjustbox}
\usepackage{booktabs}
\usepackage{tabularx}
\usepackage{hhline}
\usepackage{url}
\usepackage{hyperref}
\usepackage[square,numbers,sort&compress]{natbib}
\definecolor{revision}{RGB}{0,0,0}

\usepackage{fancyhdr}
\let\origtitle\title 
\renewcommand{\title}[1]{\lfoot{\textit{#1}}\origtitle{\textbf{#1}}}
\renewcommand{\sectionmark}[1]{\markboth {}{}}

\date{}

\title{HandyLabel: Towards Post-Processing to Real-Time Annotation Using Skeleton Based Hand Gesture Recognition}

\begin{document}
\maketitle
\thispagestyle{fancy}
\centering

\author{
Sachin Kumar Singh\footnote{ssingh@rptu.de},
Ko Watanabe\footnote{ko.watanabe@dfki.de},
Brian Moser\footnote{brian.moser@dfki.de},\\
Shoya Ishimaru\footnote{ishimaru@omu.ac.jp},
Andreas Dengel\footnote{andreas.dengel@dfki.de},
}\\
\thanks{
$^{1}$$^{5}$RPTU Kaiserslautern-Landau,
$^{2}$$^{3}$$^{5}$DFKI GmbH,\\
$^{4}$Osaka Metropolitan University,
}

\abstract{
The success of machine learning is deeply linked to the availability of high-quality training data, yet retrieving and manually labeling new data remains a time-consuming and error-prone process. Traditional annotation tools, such as \labelStudio, often require post-processing, where users label data after it has been recorded. Post-processing is highly time-consuming and labor-intensive, especially with large datasets, and may lead to erroneous annotations due to the difficulty of subjects' memory tasks when labeling cognitive activities such as emotions or comprehension levels. In this work, we introduce \appName, a real-time annotation tool that leverages hand gesture recognition to map hand signs for labeling. The application enables users to customize gesture mappings through a web-based interface, allowing for real-time annotations. To ensure the performance of \appName, we evaluate several hand gesture recognition models on an open-source hand sign (HaGRID) dataset, with and without skeleton-based preprocessing. We discovered that ResNet50 with preprocessed skeleton-based images performs an F1-score of 0.923. To validate the usability of \appName, a user study was conducted with 46 participants. The results suggest that 88.9 \% of participants preferred \appName over traditional annotation tools.
}

\section{Introduction}

\begin{figure}[t!]
  \centering
  \includegraphics[width=\textwidth]{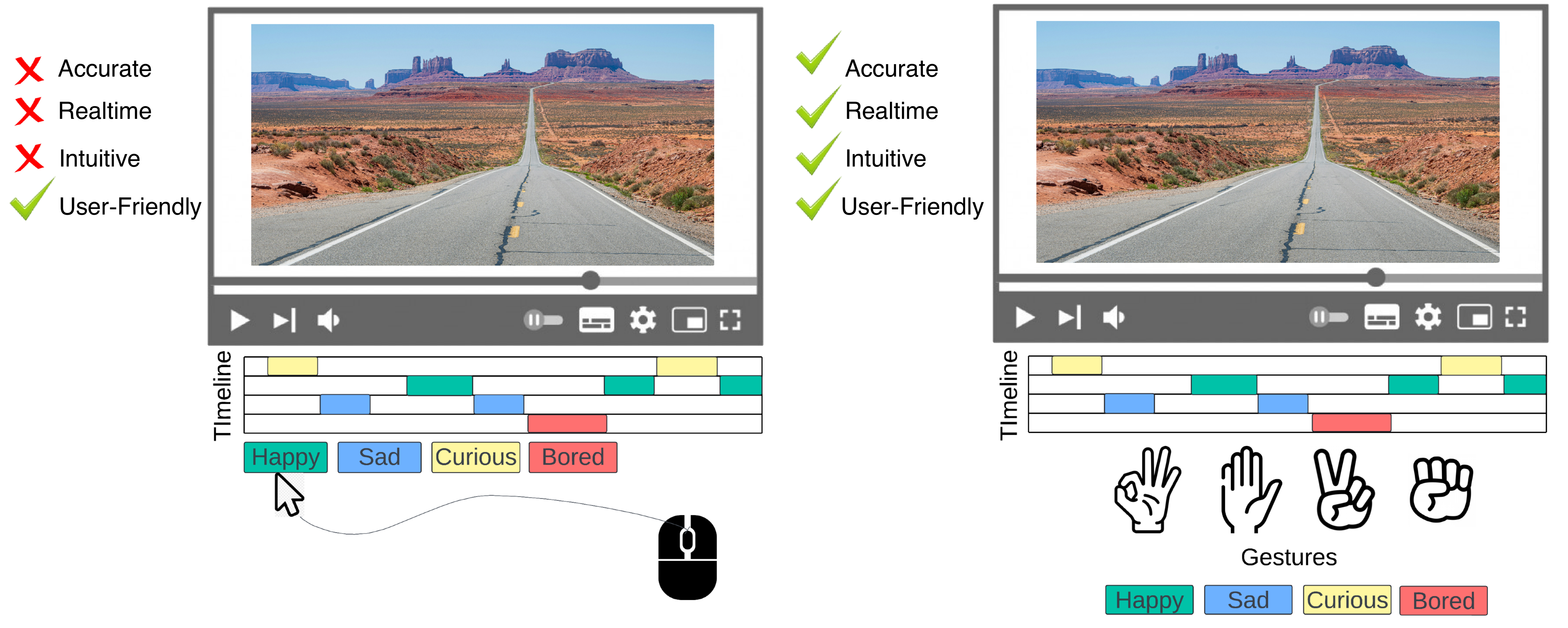}
  \caption{Overview of the \appName: Existing annotation tools perform through post-processing, where the user tries to make an annotation by recalling the target label (i.e., emotions). However, recalling might cause false memory and wrong annotation. Instead, our proposed application allows participants to make annotations during the experiment, enabling real-time precision.}
  \label{fig:teaser}
\end{figure}

The success of modern ubiquitous technology applications, or so-called \gls{ai} systems, fundamentally depends on the availability of high-quality annotated data.
From large language models~\cite{openai2025gpt5}, image generation systems~\cite{rombach2022high}, computer vision applications~\cite{ridnik2021imagenet, Zou2018}, and affective computing~\cite{cauchard2024affective, grzeszczyk2023decoding, watanabe2023engauge, watanabe2024metacognition}, the quality and quantity of training data directly determine model performance.
Pre-trained models such as BERT~\cite{Devlin2019} and GPT-5~\cite{openai2025gpt5} exemplify how large-scale, carefully annotated corpora enable breakthrough capabilities.
This dependency has elevated labeled data to the status of ``new oil'' in the \gls{ai} ecosystem, driving innovations across critical domains including autonomous driving~\cite{Caesar2020nuScenes, Sun2020Scalability}, healthcare~\cite{alabdulhafith2018customized, alwasel2025modeling}, education~\cite{sloan2021emotional, watanabe2023engauge}, and creative industries~\cite{Davis2019CreativeAI}.

Despite this critical importance, current data annotation practices remain mired in inefficient post-processing workflows.
Existing tools such as \labelStudio~\cite{LabelStudio} and CVAT~\cite{CVAT2021} require users to manually label data after recording, using GUI-based interfaces that excel in static datasets but struggle in dynamic and real-time annotations.
This limitation becomes particularly problematic as emerging applications increasingly demand immediate annotation capabilities for live data streams, interactive systems, and real-time feedback scenarios.

The human and economic costs of current annotation methods are substantial and well-documented. 
\citet{fredriksson2020data} demonstrates that expert-level image labeling requires significant time and financial investment, while \citet{rasmussen2022challenge} quantifies the temporal burden in agricultural dataset annotation.
\citet{watanabe2021discaas} further reveals that annotation tasks for online meeting participant behavior represent the most considerable workload in their system.
The fundamental bottleneck lies in operational inefficiencies: annotators must repeatedly pause and navigate through video content frame by frame, creating a time-consuming and error-prone process that constrains scalability and consistency.

To address these critical limitations, we introduce \appName, a real-time, hand gesture-based annotation system. 
As shown in \autoref{fig:teaser}, the application supports an annotation paradigm that ranges from post-processing to real-time interaction. \appName leverages hand gestures as an intuitive and natural interaction modality, enabling users to apply labels instantly during video playback without interrupting the viewing experience.
The system supports both predefined and custom gesture mappings, allowing users to mark events, emotional states, or behavioral patterns in real-time. This approach significantly reduces annotation time and improves intuitive usability..

We validate \appName's effectiveness through a comprehensive user study involving 46 participants, comparing its performance against \labelStudio across multiple dimensions, including annotation speed, user intuitiveness, and workload reduction. Our work makes three primary contributions (C1--C3):

\begin{itemize}
\item[\textbf{C1:}] \textbf{We perform significant performance on hand-gesture recognition by applying skeleton-based image preprocessing:} We compared with various deep-learning models and achieved an F1-Score of 0.92 using ResNet50.
\item[\textbf{C2:}] \textbf{We design and implement \appName, a novel real-time annotation tool:} Our application leverages skeleton-based hand gesture recognition to enable real-time data labeling from camera recordings.
\item[\textbf{C3:}] \textbf{We demonstrate that \appName significantly improves annotation speed and reduces cognitive workload:} We compared the traditional post-processing data-annotation tool to measure the usability of the system.
\end{itemize}

\section{Related Work}
In this section, we review state-of-the-art annotation platforms and gesture recognition systems, focusing on their strengths and limitations.
Building upon these foundational works, \appName integrates real-time gesture recognition capabilities with intuitive annotation workflows to address key challenges identified in prior research.

\subsection{Annotation Systems}
Label Studio~\cite{LabelStudio} is a widely used open-source annotation platform known for its versatility in handling diverse data types, including text, images, and videos.
Specifically, for video annotation, its timeline-based interface enables users to label specific segments over time.
This flexibility makes it highly adaptable to various project requirements.
However, its workflow primarily focuses on post-event annotation, requiring users to label data after recording.
This limitation makes it less ideal for applications requiring real-time interaction.
\appName bridges this gap by enabling users to annotate videos as they are recorded, streamlining the process and reducing post-event effort. 

\citet{lim2020video} proposed a hand gesture recognition system that utilizes 3D skeletal features in combination with spatial-temporal data.
Their method, designed for a single-camera setup, processes raw video sequences and uses Support Vector Machines for gesture classification.
The system achieved high accuracy across 24 gesture classes in controlled environments, eliminating the need for additional hardware like depth sensors.
This cost-effective approach makes it suitable for user-friendly applications.

GestureGAN, introduced by \citet{tang2018gesture}, employs a Generative Adversarial Network to translate hand gestures in unconstrained environments. By incorporating hand skeleton data and innovative loss functions, GestureGAN improves the quality of gesture translations while addressing challenges like "channel pollution" during training. Although GestureGAN achieved state-of-the-art results on benchmark datasets and enhanced data augmentation for classifiers, its computational demands restrict its suitability for real-time applications.

\subsection{Hand Gesture Recognition}
In this section, we explore the evolution of hand gesture recognition systems, highlighting their key contributions and limitations. These advancements have paved the way for more natural and intuitive human-computer interactions. Our review spans various methodologies, including deep learning frameworks, object detection models, and systems leveraging specialized hardware, while emphasizing how \appName addresses existing gaps through real-time adaptability and user-centric design.

Hand gesture recognition has emerged as a pivotal technology for enabling seamless interaction between humans and machines~\cite{10.1145/2069216.2069241, 10.1145/2525194.2525296}. Its applications span virtual reality, sign language interpretation, gaming, robotic control, and touchless navigation systems~\cite{sabbella2024evaluating, vaitkevivcius2019recognition, yaseen2024next, sharma2024real}. Over time, researchers have aimed to optimize accuracy, speed, and hardware simplicity, making gesture recognition systems more accessible and efficient~\cite{7301342, li2019deep}. Recent advancements in deep learning have significantly improved system performance, enabling real-time functionality even in dynamic, real-world scenarios.

\citet{10373734} developed a comprehensive gesture recognition system based on YOLO object detection frameworks. Testing several variants—YOLOv5, YOLOv6, and YOLOv8—they demonstrated the robustness of YOLOv8, achieving a remarkable 99.5\% accuracy on a customized dataset. This underscores the potential of YOLO-based models for real-time human-computer interaction. While YOLO excels in detecting objects, \appName integrates skeletal data to enhance recognition accuracy, making it more versatile for nuanced hand movements rather than general object detection.

\citet{rubin2020hand} combined Faster R-CNN with Inception V2 for dynamic gesture recognition. Their model, optimized using the ADAM algorithm, achieved an impressive precision of 99.1 and a response time of 137 milliseconds, even in challenging environments with variable lighting and complex backgrounds. Despite its strengths, the computational demands of such models can limit their practicality for real-time applications. By contrast, \appName emphasizes simplicity and real-time adaptability, leveraging skeleton-based preprocessing to achieve high accuracy while maintaining efficiency.

Generative models like GestureGAN, developed by \citet{10.1145/3240508.3240704}, have further pushed the boundaries of gesture recognition. Using a Generative Adversarial Network, GestureGAN translates hand gestures across various poses, sizes, and locations by incorporating skeleton data. While it achieves state-of-the-art performance in gesture translation and augmentation, its computational complexity makes it less suitable for real-time applications.

Systems employing traditional machine learning techniques have also contributed to the field. \citet{10.1145/2814940.2814997} designed a real-time gesture recognition system using Random Forest classifiers, achieving high accuracy in cluttered or complex backgrounds. Their system reduced posture data to lower-dimensional representations, enabling efficient recognition at over 30 frames per second. \citet{10.1145/3003733.3003746} developed a low-cost system tailored for educational applications, using a Random Forest classifier to detect 16 gestures. While accessible and practical, such systems lack the scalability and adaptability required for broader real-time applications.

Depth-based systems utilizing Kinect~\cite{10.1145/2525194.2525296} have shown remarkable effectiveness in challenging environments, achieving over 91\% accuracy for finger identification and complex gestures even under poor lighting or cluttered backgrounds. However, the reliance on specialized hardware limits their accessibility.

The MediaPipe framework~\cite{mediapipe} plays a crucial role in enhancing \appName’s capabilities. MediaPipe’s real-time pipeline tracks 21 hand landmarks, providing lightweight and accurate skeleton-based preprocessing. This approach enables \appName to adapt to variable lighting and complex environments, ensuring reliable performance across diverse scenarios. By integrating MediaPipe with transformer-based models like ViT-B16, \appName achieves a unique balance of accuracy, speed, and simplicity.

In summary, \appName builds on these innovations by integrating lightweight skeleton-based preprocessing with robust machine learning techniques. This ensures consistent performance across various environments while maintaining accessibility through standard hardware compatibility. By prioritizing real-time adaptability and user-centric design, \appName offers a practical solution that bridges existing gaps in gesture recognition technology, paving the way for more intuitive and inclusive human-computer interactions.

\section{Methodology}
In this section, we detail the design and implementation of \appName, our real-time hand gesture-based annotation tool. 
We describe the dataset selection and gesture taxonomy, the preprocessing and model training pipeline, and the system architecture that enables seamless, low-latency annotation during live video playback. 

\subsection{Dataset and Gesture Selection}
\begin{figure}[t!]
    \centering
    \begin{subfigure}[b]{0.4\textwidth}
        \centering
        \includegraphics[width=\textwidth]{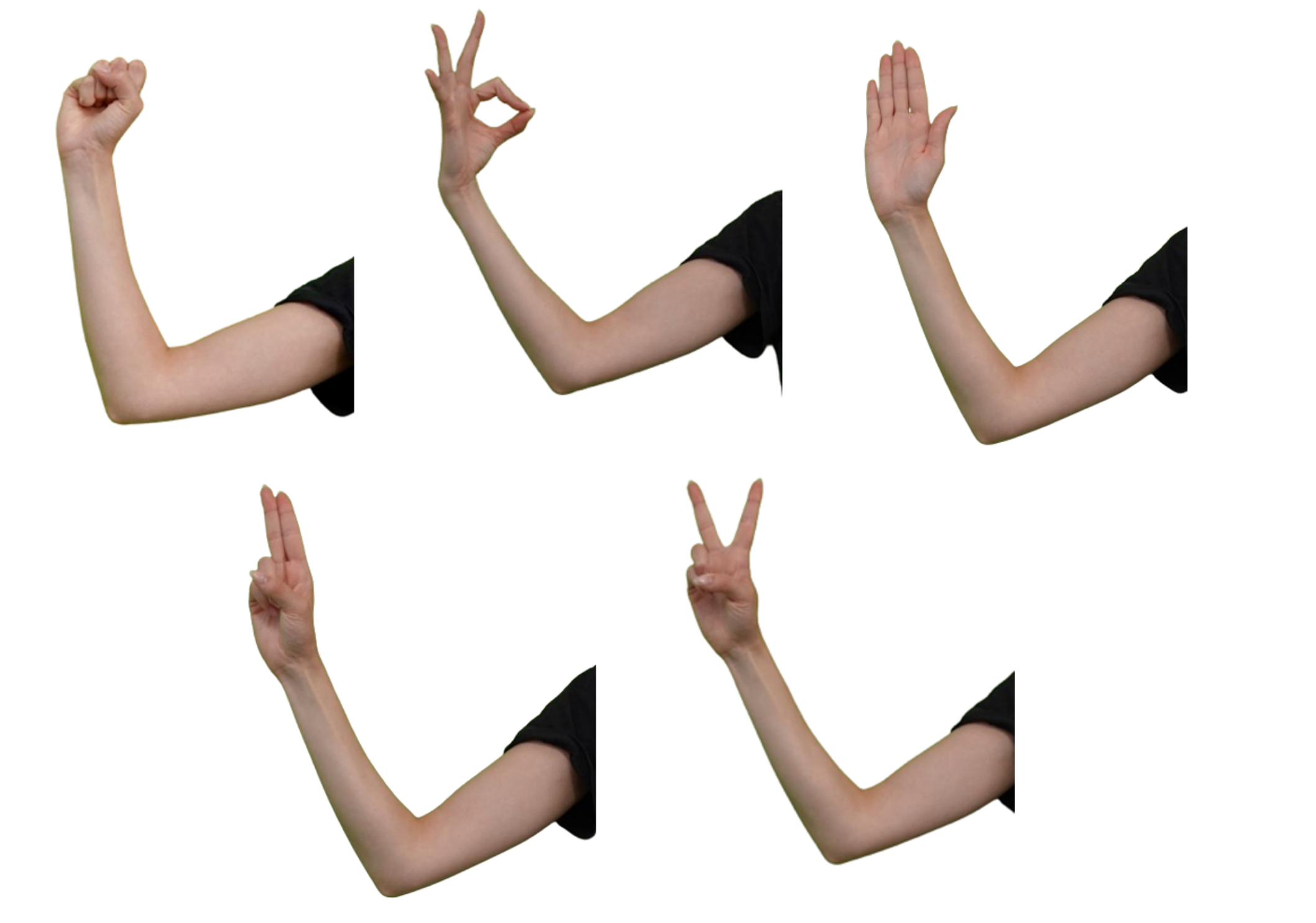}
        \caption{The five hand gestures selected for our evaluation: Fist, Ok, Stop, Two-Up, and Peace. These gestures were chosen for their distinctiveness and intuitive nature in real-time annotation scenarios.}
        \label{fig:hand_gestures}
    \end{subfigure}
    \hfill
    \begin{subfigure}[b]{0.55\textwidth}
        \centering
        \includegraphics[width=\textwidth]{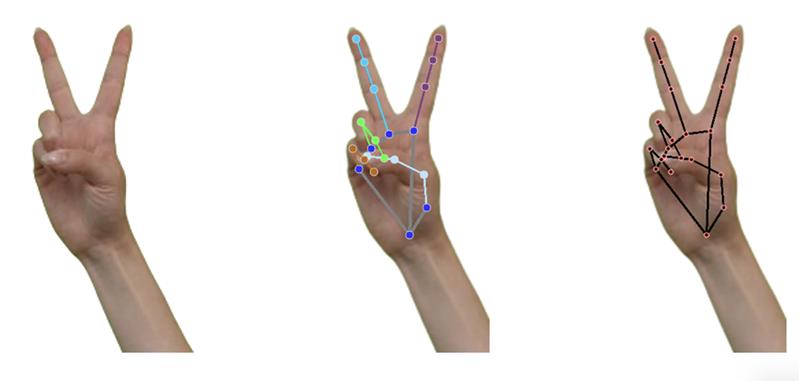}
        \caption{Data preprocessing configurations: \textbf{(left)} raw hand image, \textbf{(center)} skeleton-based preprocessing (skeleton type 1), and \textbf{(right)} show single-tone skeleton-based preprocessing (skeleton type 2) with MediaPipe~\cite{mediapipe} skeletal structure.}
        \label{fig:pre-processing}
    \end{subfigure}
    \caption{Selected hand gestures (\subref{fig:hand_gestures}) and preprocessing configurations for gesture recognition (\subref{fig:pre-processing}).}
\end{figure}

We utilize the publicly available HaGRID dataset~\cite{Kapitanov_2024_WACV}, which contains over 500,000 annotated hand images across multiple gesture classes, collected under diverse lighting and background conditions. From this comprehensive dataset, we selected five distinct gestures—\textit{Fist}, \textit{Ok}, \textit{Stop}, \textit{Two-Up}, and \textit{Peace}—based on their distinguishability, intuitive nature, and suitability for real-time annotation tasks. \autoref{fig:hand_gestures} illustrates examples of these five selected gestures.

\subsection{Data Preprocessing Strategies}
We evaluated three preprocessing approaches across all model architectures: (1) raw hand images without modification, (2) \textbf{Skeleton Type 1}: skeleton-based features extracted using the MediaPipe library as illustrated in \autoref{fig:pre-processing}, and (3) \textbf{Skeleton Type 2:} shows skeleton-based features using single-tone node and line color for skeleton.

\subsection{Model Selection and Training Protocol}
To achieve gesture recognition, we conducted a systematic comparison of multiple deep learning architectures, including ResNet18~\cite{he2016resnet18}, ResNet50~\cite{he2016resnet50}, MobileNetV3~\cite{howard2019mobilenetv3}, and ViT-B/16~\cite{dosovitskiy2021vit}.
All models were trained using cross-entropy loss with early stopping to prevent overfitting, employing a batch size of 64 and an initial learning rate of $10^{-3}$. 

Performance evaluation was conducted on a held-out test split from the HaGRID dataset to ensure unbiased assessment.
Among all evaluated architectures, ResNet50 with skeleton-based preprocessing demonstrated the highest F1 score (explained in \autoref{sec:hand_gesture}) and exhibited the most robust generalization capabilities, establishing it as the optimal backbone for \appName.

\subsection{System Deployment and Design Insights}
Beyond dataset preparation and model training, the design of \appName required careful consideration of deployment robustness, scalability, and ethical usability. This subsection summarizes key architectural and human-centered decisions that enable reliable real-time performance across diverse conditions.

\subsubsection{Robustness and Generalization Across Conditions}
The system was engineered to remain robust under uncontrolled environmental factors such as lighting variation, device quality, and user-to-camera distance. Skeleton-based preprocessing, extracted using the MediaPipe pipeline, normalizes hand appearance and decouples recognition from the raw visual background. This approach significantly reduces model sensitivity to illumination, clutter, and skin-tone diversity, enabling consistent operation across low-cost webcams and high-end imaging devices. 

In our internal evaluations, users interacted with \appName in ordinary office and home environments using standard laptop cameras. The model maintained stable detection even when users were positioned up to two meters away or partially occluded by objects such as keyboards or microphones. These observations confirm that skeletal representations effectively mitigate domain shift between laboratory and real-world deployments—a critical requirement for behavioral and affective annotation scenarios.

\subsubsection{Scalability and Real-Time Responsiveness}
From the outset, scalability was prioritized at two complementary levels: computational and cognitive. 
Computationally, \appName employs a modular client–server design in which the browser-based frontend captures frames and transmits skeleton keypoints to a lightweight Python backend via asynchronous sockets. This architecture maintains low latency and supports multiple concurrent users. In a local network configuration, end-to-end gesture recognition latency averaged under 100~ms, well below the perceptual threshold for real-time responsiveness.

Cognitive scalability focuses on reducing user workload during extended annotation sessions. By mapping intuitive gestures to semantically meaningful labels, \appName minimizes mental overhead compared to traditional keyboard-based interfaces. Pilot deployments in mock lecture and driving scenarios showed that participants could sustain accurate annotations for more than 30~minutes without notable fatigue. These findings underscore the system’s suitability for long-duration studies such as classroom monitoring or simulator-based experiments.

\subsubsection{Practical Examples of Deployment Contexts}
To illustrate its versatility, we briefly discuss three example contexts where the same architecture can be applied with minimal adaptation.

\textbf{Real-Time Communication Add-On:} As a plug-in for video-conferencing platforms such as Zoom or Microsoft Teams, \appName could interpret gestures like “stop,” “ok,” or “thumbs up” as meeting control commands or engagement signals. Aggregating these gestures across participants would allow automated detection of collective agreement or confusion moments, enhancing accessibility and collaboration.

\textbf{Driver–Vehicle Interaction:} In automotive research, \appName could capture non-verbal driver feedback in real time. For example, repeated “stop” gestures may indicate cognitive overload, prompting adaptive vehicle interfaces to reduce notifications or suggest a rest break. Integration with gaze and steering sensors would enable richer multimodal behavioral datasets for safe and adaptive driving systems.

\textbf{Audience Feedback Analytics:} In media or educational settings, multiple users could use simple gestures to express emotional responses—such as “fist” for excitement or “peace” for surprise—while watching a video or presentation. Aggregated recognition data can generate engagement heatmaps or temporal emotion curves, offering creators and educators immediate feedback without interrupting the experience.

\subsubsection{System Integration and Engineering Reflection}
From an engineering standpoint, using a web-based frontend with a lightweight Python backend ensures cross-platform compatibility. The Flask server hosts the trained ResNet50 model, while the frontend implements real-time visualization through WebSockets. MediaPipe’s 21-landmark extraction minimizes data size, allowing transmission of only keypoint coordinates instead of full video frames. This design reduces bandwidth requirements and preserves user privacy. The architecture also supports future extensions, such as substituting the backend model with a transformer-based recognizer or integrating cloud-based logging for large-scale studies.

\subsubsection{Ethical and Human-Centered Design}
Privacy and transparency guided all deployment decisions. \appName avoids storing raw video streams; only skeletal coordinates and derived annotation metadata are saved locally or transmitted for analysis. Users maintain full control over gesture–label mappings, accommodating cultural or physical differences in gesture expression. This inclusive approach aligns with human-centered AI principles, ensuring the system functions not merely as a technical demonstrator but as a privacy-conscious and adaptable research tool.

\subsubsection{System Summary}
By combining robustness, real-time responsiveness, scalability, and ethical design, \appName demonstrates how a single skeleton-based framework can serve both as an efficient annotation tool and as a foundation for multimodal behavioral computing applications. These design insights collectively strengthen the system’s readiness for deployment across diverse experimental and applied contexts.

\section{Result and Discussion}
This section presents the comprehensive evaluation of \appName across two critical dimensions: technical performance and user experience. 
We first examine the hand gesture recognition capabilities through systematic model comparisons, followed by a detailed usability study that validates the tool's practical effectiveness in real-world annotation scenarios.

\subsection{Hand Gesture Recognition Model Selection}
\label{sec:hand_gesture}
\autoref{tab:model_config} shows the F1 scores of various models with and without skeleton-based preprocessing on the HaGRID dataset. 
As a result, we found that using Skelton Type 1 mediapipe processing performed the best with ResNet50 on the F1 score of 0.923 for recognizing the target five-class hand gestures.

According to the result, in our application, \appName uses a ResNet50 backbone and MediaPipe skeleton keypoints for real-time gesture recognition. 
The model works on a standard laptop webcam and does not require special hardware or calibration. 
Users can assign labels to gestures and see live recognition results with timestamps in the interface. 
The system is designed for responsiveness and works reliably under normal indoor conditions. 
Visual feedback helps users adjust their gestures as needed. The modular design allows easy adaptation to other applications.

\begin{table}[t!]
\centering
\renewcommand{\arraystretch}{1.0}
\caption{Comparison of F1 score for different model architectures and different preprocess conditions across configurations on the HaGRID test dataset.}
\label{tab:model_config}
\begin{tabular}{l|c|c|c}
    \hline
                   & \multicolumn{3}{c}{\textbf{Preprocess Condition}} \\
    \cline{2-4}
    \textbf{Model} & \textbf{Baseline} & \textbf{Skeleton Type 1} & \textbf{Skeleton Type 2} \\
    \hline
    ResNet18 & 0.8960 & \cellcolor{blue!15}0.9171 & 0.8092 \\
    MobileNetV3\_S & 0.8248 & \cellcolor{blue!15}0.9143 & 0.7649 \\
    MobileNetV3\_L & \cellcolor{blue!10}0.8253 & 0.8245 & 0.6790 \\
    ResNeXt50 & \cellcolor{blue!10}0.8902 & 0.8757 & 0.8740 \\
    ResNet50 & 0.8862 & \cellcolor{blue!30}0.9230 & 0.8765 \\
    ViTB16 & \cellcolor{blue!10}0.8948 & 0.8222 & 0.7829 \\
    \hline
\end{tabular}
\end{table}

\subsection{END-USER APPLICATION: \appName}
\label{section:application_scenario}
\begin{figure}[t!]
    \centering
    \includegraphics[width=\textwidth]{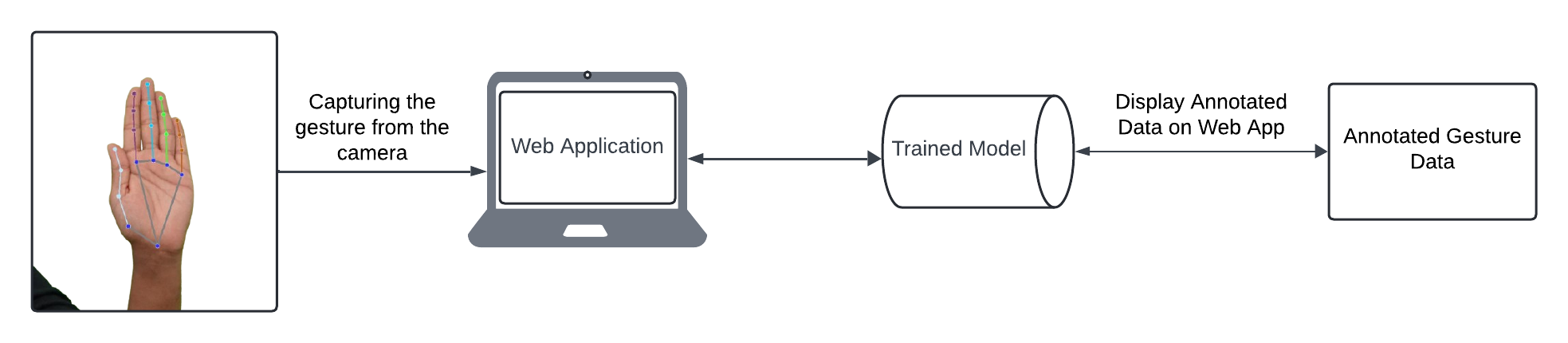}
    \hfill
    \caption{System overview. The workflow begins with real-time hand gesture capture via a camera, followed by immediate processing in the web application. Captured frames are transmitted to a backend server, where a skeleton-based ResNet50 model recognizes the gestures and generates corresponding annotations. These annotated results are instantly visualized in the web interface, enabling users to perform fast, accurate, and intuitive data labeling with minimal latency.}
    \label{fig:overview_app}
\end{figure}

\appName is designed as a real-time hand gesture recognition system that enables intuitive, instantaneous data annotation. 
\textcolor{revision}{Our application focuses mainly on annotations better suited for real-time events, such as cognitive or emotional states for each individual. The system will avoid recall-based annotations after the event, which might lead to false memories and incorrect annotations.}
\autoref{fig:overview_app} illustrates the system workflow and the web application user-interface.
The complete implementation is available as open-source software~\footnote{\textbf{APPLICATION LINK: \url{https://anonymous.4open.science/r/handylabel}}}.

\subsubsection{System Architecture and Implementation}
The system leverages hand gestures as a natural interaction modality, enabling users to annotate real-time events, emotional states, or behavioral patterns without interrupting their primary activities.
The architecture prioritizes three key objectives: ease of use, high recognition accuracy, and minimal latency for real-time applications. 
The system is designed to accommodate users across the expertise spectrum, from novices to experienced annotators, ensuring both simplicity and powerful functionality.

\begin{figure}[t!]
    \centering
    \begin{subfigure}[b]{0.6\textwidth}
        \centering
        \includegraphics[width=\textwidth]{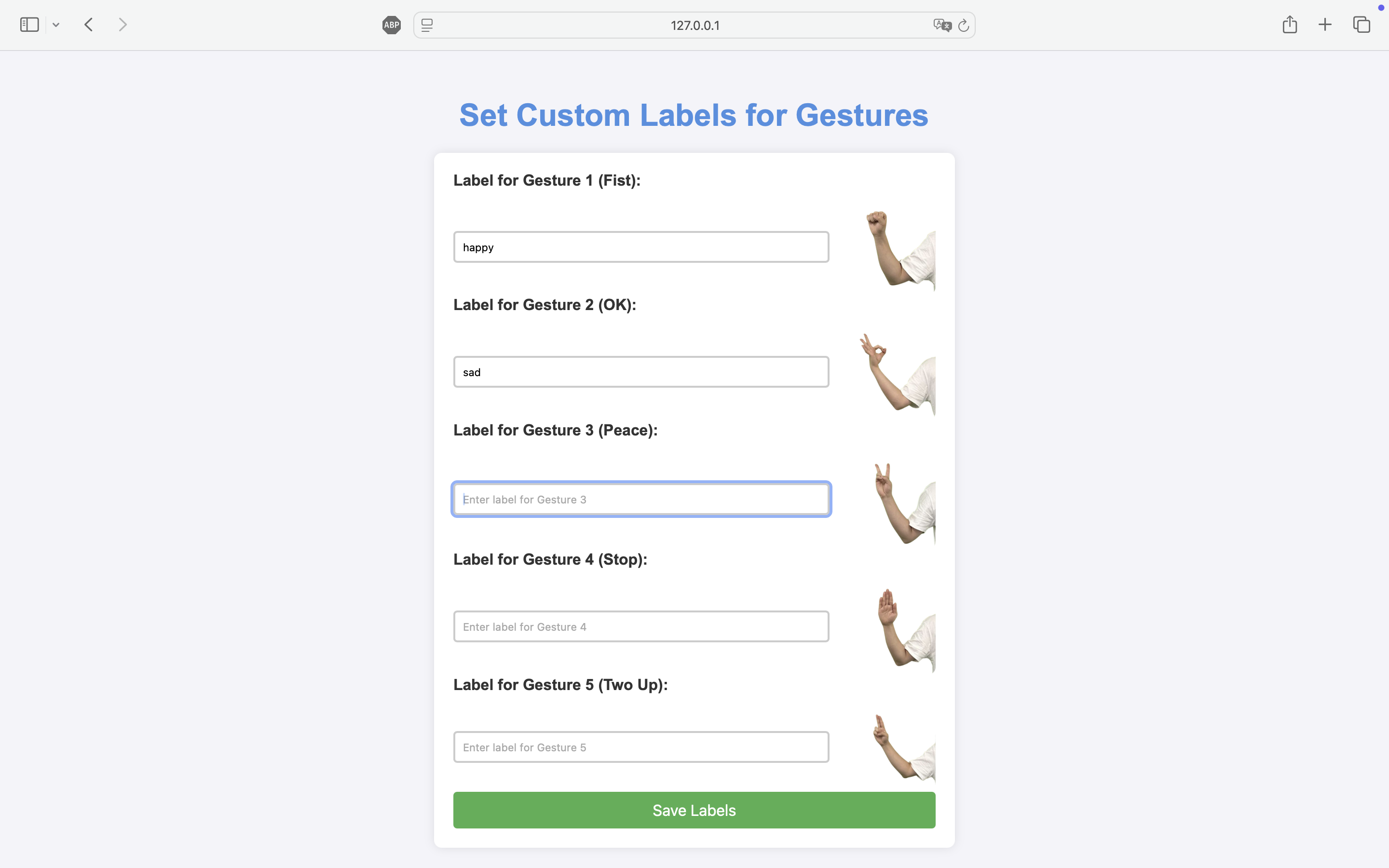}
        \caption{Step 1: Label selection phase.}
        \label{fig:app_stage1}
    \end{subfigure}
    \begin{subfigure}[b]{0.6\textwidth}
        \centering
        \includegraphics[width=\textwidth]{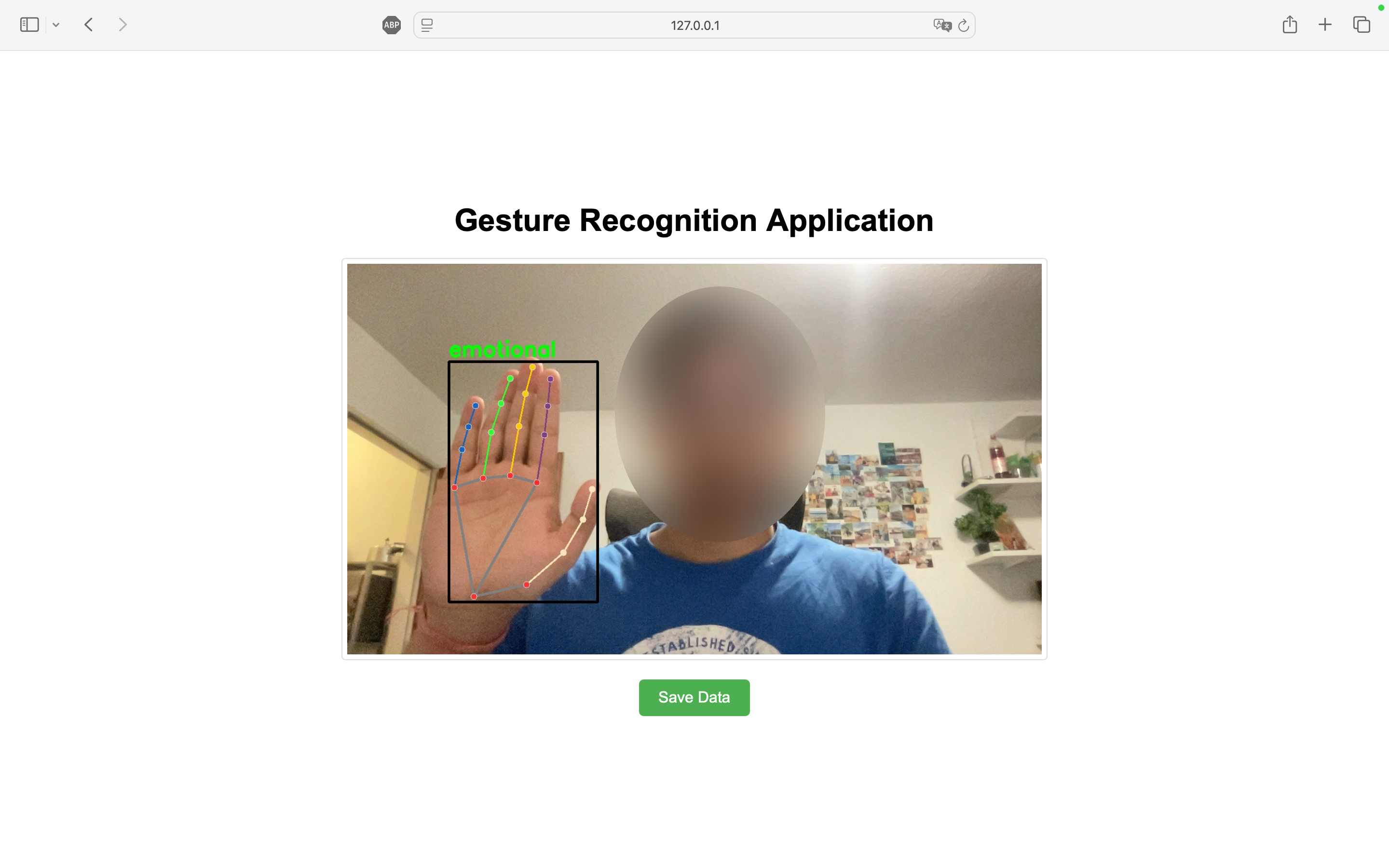}
        \caption{Step 2: Annotation phase.}
        \label{fig:app_stage2}
    \end{subfigure}
    \begin{subfigure}[b]{0.6\textwidth}
        \centering
        \includegraphics[width=\textwidth]{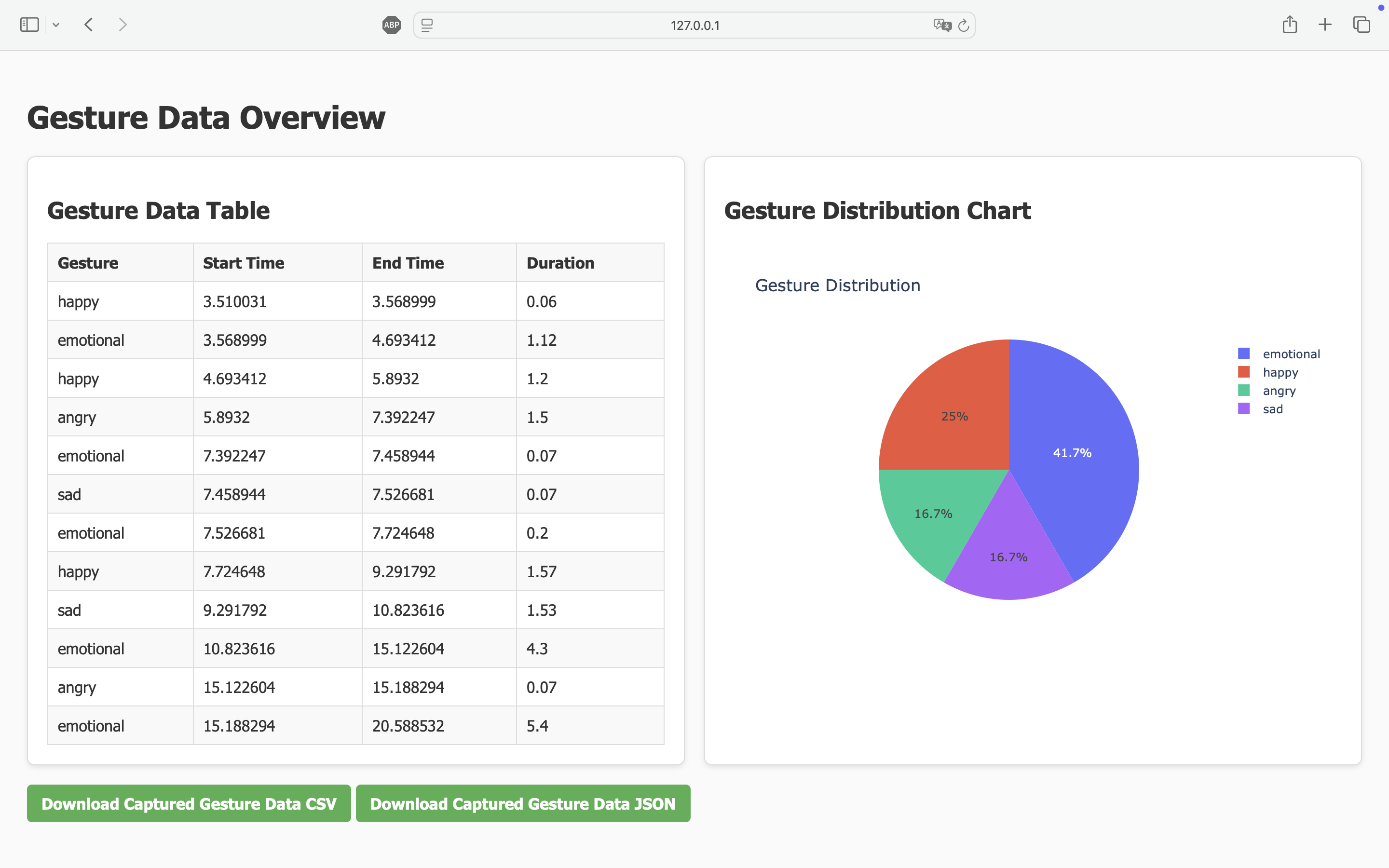}
        \caption{Step 3: Data visualization and export phase.}
        \label{fig:app_stage3}
    \end{subfigure}
    \caption{\appName workflow: (\subref{fig:app_stage1}) Users select one-to-one relationship of the annotation label and hand sign, (\subref{fig:app_stage2}) Camera recording starts and hand sign recognition will be running in the backend and when model recognize the selected hand gestures, then the timestamp and the label will be stored as a log. (\subref{fig:app_stage3}) Lastly, after user stop recording, the dashboard shows the datetime/timestamp and the duration of each annotation. The data can be exported as a CSV file.}
    \label{fig:application_workflow}
\end{figure}

The user interface provides an intuitive workflow for gesture-based annotation. \autoref{fig:application_workflow} illustrates the application workflow. 
Upon system initialization, users can define custom labels for up to five gestures, enabling adaptation to diverse use cases such as emotional annotation during media consumption or real-time feedback collection during presentations.
\appName employs MediaPipe~\cite{mediapipe} to track 21 hand keypoints, which significantly enhances gesture recognition accuracy while providing real-time visual feedback to users. 
This skeleton-based approach ensures precise, low-latency annotations that consistently outperform models relying solely on raw hand images.

The system architecture employs a client-server model with a Flask-based backend that handles computationally intensive gesture recognition tasks. 
The frontend captures camera frames in real-time and transmits them to the backend for processing using models trained on the HaGRID dataset~\cite{Kapitanov_2024_WACV}. 
The skeleton-based preprocessing pipeline ensures robust performance by focusing on essential hand movements, maintaining high accuracy even in visually complex environments with varying lighting conditions and backgrounds.

A core innovation of \appName is its real-time activity annotation capability, which enables continuous logging of recognized gestures with millisecond-precision timestamps. 
This temporal precision is crucial for applications requiring detailed behavioral analysis, such as tracking emotional trajectories during media consumption or monitoring user engagement patterns during presentations.

The system provides flexible data management options, allowing users to save recorded gestures and their corresponding timestamps immediately after recognition or accumulate data for batch processing. 
This dual approach accommodates both immediate analysis needs and comprehensive post-event evaluation scenarios.

\appName exports annotation data in CSV format, ensuring compatibility with standard data analysis tools and workflows. 
The system includes comprehensive visualization capabilities that transform raw annotation data into actionable insights through multiple representation formats:

\begin{itemize}
    \item \textbf{Timeline visualizations} that display gesture occurrences chronologically, enabling temporal pattern analysis.
    \item \textbf{Frequency breakdowns} that quantify gesture usage patterns and identify dominant interaction behaviors.
    \item \textbf{Pie charts} that illustrate the distribution of gesture types throughout recording sessions.
\end{itemize}

These visualization tools provide researchers and practitioners with immediate insights into user behavior patterns, emotional responses, and interaction dynamics, supporting applications in affective computing, user experience research, and real-time feedback systems.

\subsubsection{Usability Study and Experiment Setup}
To assess the perception and usability of \appName, we conducted a comprehensive user study comparing it with the established annotation tool \labelStudio. 
The study's primary goals were to evaluate the intuitiveness, ease of use, and suitability of both tools for annotation tasks.

\begin{figure}[t!]
  \centering
  \includegraphics[width=\linewidth]{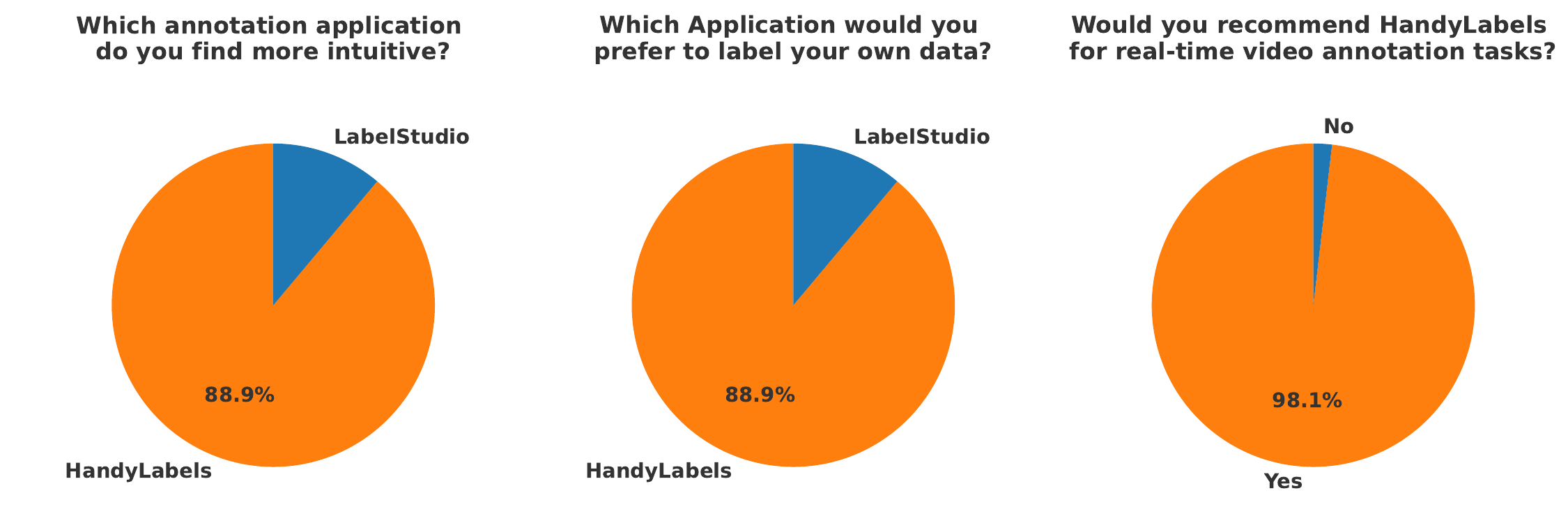}
      \caption{User preferences for \appName compared to \labelStudio. The first pie chart illustrates that the vast majority of users found \appName more intuitive than \labelStudio, with only a small percentage favoring \labelStudio. The second chart reflects user preferences for data annotation, with 88.9\% of participants indicating they would rather use \appName for annotation tasks. The third chart shows that the majority would recommend \appName, highlighting a strong overall preference for \appName.}
      \label{fig:pie1}
  \end{figure}

To assess the perception and usability of \appName, we compared a comprehensive user study with the established annotation tool \labelStudio. 
Label Studio was chosen as the baseline due to its widespread adoption in research and industry, its comparable open-source architecture, and its support for multi-modal annotation workflows. 
This makes it a representative and fair comparison point for evaluating \appName. 
Future work will expand this evaluation to include standardized usability questionnaires such as the System Usability Scale (SUS) and comparisons with additional commercial annotation platforms.
The study's primary goals were to evaluate the intuitiveness, ease of use, and suitability of both tools for annotation tasks.

\textcolor{revision}{
In order to compare the usability of \appName and \labelStudio, we prepare the annotation task of emotion labeling during video watching.
In the sample, emotion annotation is done against the same video and same emotion labels.
Participants will be shown the full-procedure of how each \appName and \labelStudio annotation is done.
}

The study involved 46 participants.
\textcolor{revision}{Demographic information about the participants is summarized as follows: Age of $27.45 \pm 4.37$ years old. Gender of male (37), female (9), and prefer not to say (1). Nationality of Indian (33), German (6), Japanese (3), Pakistani (1), Egyptian (1), Russian (1), Iranian (1), and Ecuadorian (1). Occupation of students (31), and workers (16)}.
Our user study collects data using a Microsoft Forms survey~\footnote{Microsoft Form Survey: \textcolor{revision}{\url{https://forms.office.com/r/SD7Bv9EMcR}}}.
Each participant was presented with videos demonstrating the labeling process using both \appName and \labelStudio. 
After watching the videos, participants answered questions assessing their experience with each tool.
This included how intuitive they found each tool, the amount of effort required for setup, and which tool they preferred for labeling tasks.

\subsubsection{User Study Results}
\begin{figure}[t!]
  \centering
  \includegraphics[width=0.7\textwidth]{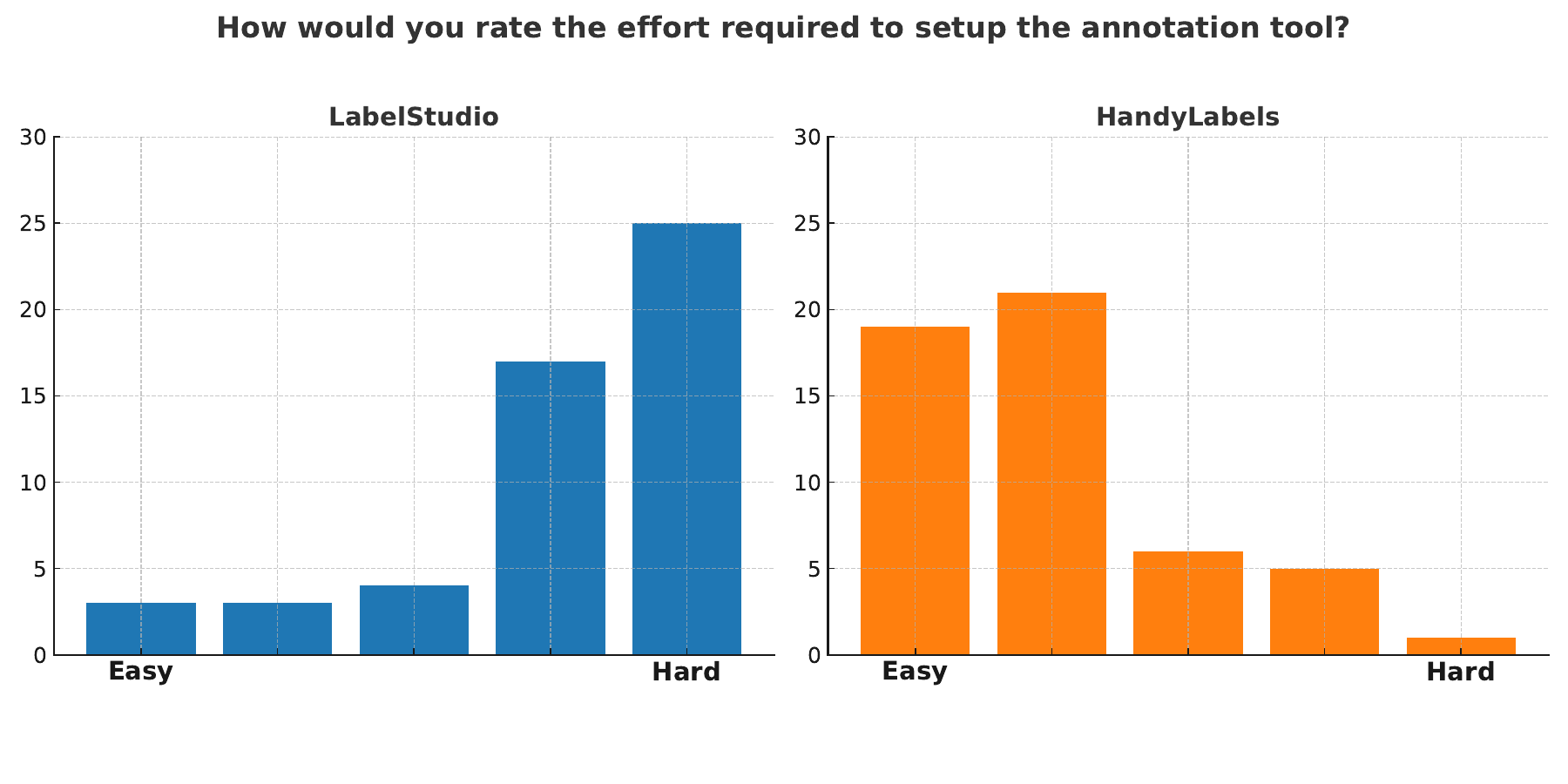}
      \caption{Comparison of user ratings for setup effort between \appName and \labelStudio. The left bar chart indicates that most participants found \labelStudio more challenging to set up, with the majority assigning it a difficulty rating of 4 or 5 (on a scale of 1 to 5, where five is the most difficult). In contrast, the right bar chart demonstrates that \appName is much easier to set up, with most participants rating the effort as 1 or 2, indicating its user-friendly setup process.}
      \label{fig:bargraph1}
\end{figure}

\appName was rated as the more intuitive tool compared to \labelStudio, with users noting its real-time gesture recognition capabilities and interactive, user-friendly interface. 
Also, participants stated that they would recommend \appName for real-time video annotation tasks, reinforcing its perceived effectiveness and ease of use.
All these three results are visualized in \autoref{fig:pie1}. Finally, participants found \appName significantly easier to set up, rating it 2.07 on a 5-point scale (where 1 = Easy and 5 = Hard), compared to 4.22 for \labelStudio, which was perceived as more manual and labor-intensive, as shown in \autoref{fig:bargraph1}.

\section{Limitation and Future Work}
The experiments conducted in this study highlight the transformative impact of skeleton-based preprocessing on hand gesture recognition systems. 
By leveraging skeletal data, the system achieved notable improvements in accuracy and robustness across diverse conditions, including varied backgrounds, devices, and distances. 
This approach allows models to isolate critical hand movements, enhancing recognition performance even in visually complex or dynamic environments.

\textbf{Dependence on Background Complexity:} A significant limitation observed in this study is the sensitivity of gesture recognition models to background complexity. Models relying solely on raw hand images showed sharp declines in accuracy when tested against intricate backgrounds. For instance, ResNet18, while effective under simple conditions, struggled as background complexity increased. However, skeleton-based preprocessing successfully mitigated this challenge, enabling the system to focus on essential hand features and maintain consistent performance across scenarios. Despite this improvement, achieving optimal performance in highly cluttered or rapidly changing environments remains an area for further exploration.

\textbf{Hardware Dependency:} Device quality emerged as another critical factor influencing recognition accuracy.
Advanced hardware, such as the iPhone 14 Pro Max, consistently outperformed lower-spec devices, owing to its superior camera capabilities. 
Nevertheless, skeleton-based preprocessing helped reduce performance disparities across devices, allowing even mid-range hardware, like the Sony X1, to deliver competitive results.
While this flexibility makes the system broadly accessible, optimizing the tool for resource-constrained environments or older hardware remains a challenge.

\textbf{Impact of Distance:} The experiments also revealed that distance from the camera affects recognition accuracy, especially for models without preprocessing. As the user moved further away from the camera, raw image-based models experienced a steep drop in performance. Skeleton-based preprocessing, however, stabilized accuracy over longer distances. For example, ViT-B16 demonstrated reliable performance up to three meters, proving its suitability for use in large spaces like classrooms, conference rooms, and open areas. Expanding the system's range while maintaining accuracy under extreme conditions will be a critical area for future work.

\textcolor{revision}{
\textbf{Number of Gestures Available for Annotation:} The number of annotations is limited. The system can handle only five hand signs we have selected as for the high recognition accuracy. The limit in the number of gestures limit the number of annotations possible. Also, this limit the users to select the natural gesture for each annotation combination. Hence, our future work will be to enable more gestures for annotation, allowing each user to select a natural gesture for annotation combinations. 
}

\section{Conclusion} 
We introduced \appName, a real-time hand gesture recognition tool that streamlines data annotation through live gesture-based inputs. We confirm that the ResNet50 transformer model performs an F1 score of 0.923 when applying skeleton-based preprocessing. Using the best performing model, we implemented \appName and conducted a user study with 46 participants, comparing it against \labelStudio. We confirmed that 88.9\% of users preferred \appName as a better labeling tool. Our results establish \appName as a practical, user-friendly solution for efficient annotation tasks.

\section*{Acknowledgment}
This work was supported by the Osaka Metropolitan University Strategic Research Promotion Program.

\bibliographystyle{plainnat}
\bibliography{main}
\end{document}